# Decentralized Blockchain-based model for Edge Computing

Mabrook S. Al-Rakhami[1,2], Abdu Gumaei[2], Sk. Md. Mizanur Rahman[3] and Atif Al-Amri[1,4]

[1]Research Chair of Pervasive and Mobile Computing; King Saud University, Riyadh 11543, Saudi Arabia
[2]Information Systems Department, College of Computer and Information Sciences, King Saud University, Riyadh 11543, Saudi Arabia
[3]Information and Communication Engineering Technology (ICET), School of Engineering Technology and Applied Science (SETAS), Centennial College, Toronto, ON M1G 3T8, Canada
[4]Software Engineering Department, College of Computer and Information Sciences, King Saud University, Riyadh 11543, Saudi Arabia
malrakhami@ksu.edu.sa, abdugomai@gmail.com, SRahman@centennialcollege.ca, atif@ksu.edu.sa

*Abstract*— **Blockchain technology is among the fastest-growing technologies in the world today. It has been adopted in diverse areas but mostly in financial systems, such as Bitcoin cryptocurrency. Therefore, it is a niche that has attracted interest from researchers from various fields, including computer science. Other areas where Blockchain is being embraced are the Smart Grid and Internet of Things (IoT) technologies, among others. While it is all good and improving many areas of applications, Blockchain still has some shortcomings. For example, it is not designed for high scalability when accommodating normal transactions. On the other hand, a parallel technology that has diverse applications in distributed networks better known as edge computing has emerged. Its main advantage is that it increases the speed of pf processes within those networks. However, like Blockchain, edge computing has its shortcomings. Its security systems and management systems have been found to be wanting. Hence the idea to integrate the two technologies and take advantage of their strengths. A blend of the two would lead to advanced network servers, huge data storage, and heightened security in transactions. However, this integration will best happen when some measures are taken. For example, there is a need to address scalability, resource management satisfactorily, and the security of the systems. To solve the integration problem, a decentralized Blockchain-based model of Edge computing is proposed in this paper.**

*Keywords— Blockchain, Edge Computing, Decentralized model.*

## I. INTRODUCTION

With the emergence of the Internet of Things technology [1], many research fields have cropped up in the field of technology. In order to realize the potential of this technology, however, there is a great need to address the challenges that it has brought with it. In six years' time, the volume of revenue that Blockchain is estimated to raise annually will rise beyond US$ 19.9. This projected growth has attracted many industry players who are now keen on ditching centralized models and adopting this new decentralized model [2]. Blockchain technology was invented on the same grounds as Bitcoin in order to address the concerns of transacting money within a centralized model [3]. The main advantage of Blockchain is that it allows a high level of transparency. This allows the transacting parties to verify the asset ownership and also sign the agreement within the platform with ease. In addition, it is decentralized and secure. However, the outstanding challenge is on scalability, which means that its applications are limited [4].

A similar technology that has become popular is edge computing. It is more of an extended function of cloud computing. As such, both technologies allow numerous devices to be in use within the infrastructure of the network. Edge computing has an advantage over cloud computing as it is faster in terms of storage and has greater computing power. These and other features make edge computing better in terms of the Quality of Service (QoS) that it offers to the devices at the edge. Although it has its good side, edge computing has shortcomings, including questionable privacy and security measures. The shortcomings are due to the fact that there is a combination of service migration and heterogeneous nodes running across the given nodes [5], [6]. Currently, combining both Blockchain technology and edge computing technology gives a system that has greater and more efficient features that include security and better management. The new system also has an assurance of integrity and validity, among other features [7]. The new system also effectively eliminates the burden of large storage as the resources being transacted through the system is distributed and not merely stored [8]. However, with all the advantages that come with integrating the systems, there are challenges that need to be addressed. These include privacy and security concerns, which are especially posed because there are services that are outsourced to support operations at the edge of the networks. As such, further research is needed to determine how such issues can practically be addressed in a satisfactory manner.

In this paper, we propose the development of a novel model that will guide the integration of both blockchain technology and edge computing. The proposed model is aimed at creating reliable and efficient access to the networks by numerous devices. It also aims at distributing the storage of resources at the edge. The outcome is the provision of servers at a large scale, increasing data storage and its validity. More importantly, to improve the security of the system.

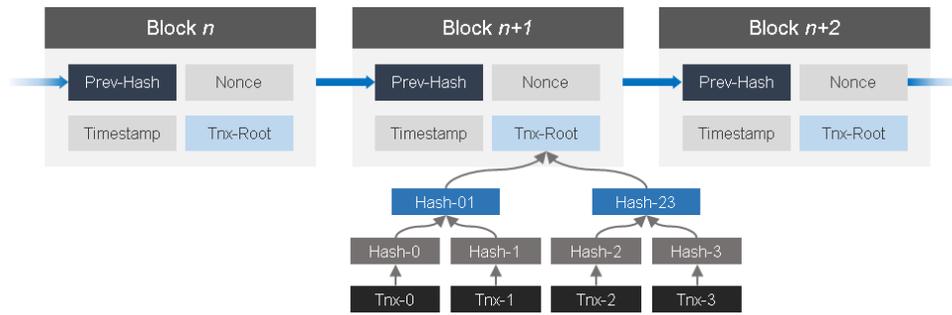

Fig. 1. Example of who blockahin handle transactions

The parts of this paper include the following: Section II provides a detailed summary of blockchain technology and edge computing; Section III provides the rationale of integrating the two technologies; Section IV provides the intricate details of the proposed model; Section V Presents the outcome of the experiment; and finally ssection VI details the conclusion.

## II. BACKGROUND

### A. Blockchain Technology

Blockchain technology is basically a distributed system of recording information in a manner that makes it impossible to change, hack, or cheat. It is a digital ledger of transactions that is duplicated and distributed across an entire network of a connected computer system on a blockchain. The system supports bitcoins and other cryptocurrencies that are transacted within the blockchain technology. The system has a high degree of transparency as all members involved in a transaction usually have a duplicate o the transaction being undertaken [9]. It is a technology which allows effective and highly secure transmission of digital information. It has roots in Bitcoin but it has grown to find applications in numerous fields. In Fig. 1, the transmission of digital information from the source to the receiver along a blockchain system is illustrated, together with actions that occur along the way.

Layers: comprehending Blockchain technology is more effective when dissected into security and the economics layers. Additionally, this breakdown offers background into the developments that have been made to the system over the past. These particular layers include the following: data, network, consensus, ledger topography, the incentive contract, and finally, the application layers [10].

- Data layer: this layer dissects data from different applications through transactions and the blocks. Along with the blockchain system, there forms a series of blocks to indicate that transactions have taken place within the network. The blocks work in a way that when one is formed, it allows the one that follows it to be fully formed. Metadata that includes the hash of the preceding and following blocks is recorded in detail the timestamp of the time the block used to form.

- Transaction layer: once a transaction is generated and verified, it is moved to the following nodes. As a result, the process details transaction using the blockchain technology. The blockchain network involves peers who are connected together in the network.

- Consensus layer: it consists of an algorithm that verifies agreement between two unknown nodes within the blockchain network. It is made up of protocols that interpret public ledger based on the format. The protocols identify consensus among ledgers and verify other blocks that are new in the ledger.

There are two major consensus algorithms. These are Proof-of-Work (PoW) and Proof-of-Stake (PoS). They are described below.

- o Proof-of-work (PoW): this is designed to ensure that security protocols of the system are not broken. Such security threats stopped by PoW include the Distributed Denial of Service (DDOS), where attackers can cram the system with false requests leading to a denial of service for genuine users. The assumption is that any nodes which are an attack are less powerful compared to the network [9]. Transactions are verified through mining, as well as the generation of new blocks.

- o Proof-of-Stake (PoS): this technique is also used in realizing consensus among different nodes in the system. The algorithm uses a random selection method to choose the creator of the subsequent blocks in the network. The other factor considered in the consensus is the stake, which may refer to wealth or age. For instance, the algorithm will give preference to the nodes bearing coins with the highest age [11]. The basis of this method is that not more than a third of the nodes in the network can be compromised. As such, nodes function together in coming to a consensus regarding the given state of the blockchain network.

- o The Ledger topography layer: its main function is to store data that has been verified by the consensus layer. The need to improve on this layer to increase its scalability has been observed. Hence, there has been the development of sidechains that allow lowered decentralization [12].

- Incentive layer: also called the mining layer, it ensures that consensus in the blockchain network exists, and the addition of new blocks to the ledger takes place. This is achieved by the use of incentives that encourage nodes to take part in the verification of data.

- Contract layer: its core function is to combine features that can be programmed. It uses the computer nodes in processing agreements that are signed between consenting nodes. These contracts are then moved to a proof public database where they cannot be distorted. The layer is fundamental in ensuring that third parties do not have their way and that only verified transactions are bound to occur upon successfully meeting conditions.
- Application layer: it ensures that services reach the end-user in the network. The layer is more involved in the application of Blockchain in various fields such as finance.

*Characterization:* Blockchain technology is unique in a number of ways. It supports a distributed transfer and access to information. It supports the storage of critical digital information in pubic databases that are proof of any interference from third parties. This digital information known as blocks is shared among members of a given chain or network. The transfer of this information is a transparent process where each member can track the progress. There is great room for research on this decentralized and also transparent system, which could lead to possible applications in diverse fields.

*B. Edge Computing*

Edge computing technology brings distributed devices within a network together. The technology, also called fog computing, eliminates the need for a central server by decentralizing the function of computer request processing. In the end, the need to use central services from the cloud is done away with, thereby increasing the speed of information transfer across the network. The speed and high-performance ideas can be implemented in the field, such as the Internet of Things (IoT) [13].

*Architecture:* the idea of edge computing is founded to reduce processing speed and increase the efficiency of the system, the network, and the end devices. The architecture consists of three distinct levels:

- End devise or (frontend): here, the devices that produce and consume data are connected within the network to interact with each other. Examples include sensors and actuators, among others.
- Edge server or (near-end): the servers support the connection between devices at the edge of the connection and the other areas of the network. The edge server's key function is to increase efficiency in connectivity through caching, improved data protection, and processing.
- Core cloud (far-end): This is situated at the far end and processes requests coming from the edge server within the network.

*Characterization:* Computing occurs at the edge, closer to the end devices. This reduces the distance that digital information has to travel. As a result, speed is increased, and so is efficiency. As a result, the cost of Computing is significantly lowered, and the experience of the system user is enhanced. The following are features that make edge computing technology fit for distributed systems.

- Intelligence and the control of processes: The edge computing technology is decentralized. As such, it becomes a great alternative for processing and handling large quantities of data. This is unlike operating on a central server where system failure could result in a total breakdown of the system.
- Less concentration of data and assured privacy: Edge computing technology is distributed. As such, it makes it more secure and improves its privacy. Besides, it can apply cloudlets that facilitate the transfer of more sensitive information within the network.
- Heterogeneous factor and scalability: edge computing technology solves the concern of scalability. This means that its applications can range from small scale to large scale. Besides, it has been found to work best with heterogeneous systems. Besides, it serves devices based on their geographic location. This is done in place of having to transfer data to the cloud for processing and computation.

III. MOTIVATIONS

This part describes the intricacies of merging both edge computing and blockchain technologies. We start by explaining the rationale behind the idea.

In order to understand the integration of the two technologies, it is appropriate that we explore both technologies. To begin with, Blockchain technology has a very secure infrastructure in place. Also, it is distributed or decentralized, which means that there is no need for a central server. Therefore, the digital information of users, transferred within a distributed ledger is secure from third parties. More importantly, Blockchain features that include ledger topology, incentive layer, and consensus algorithm are key in securing data and decentralizing processes. The edge computing, on its part, makes it possible for devices to operate at the edge of the computing network, therefore reducing the route of data processing and increasing the speed and efficiency of the processes. The fundamental benefit of edge computing technology is that it has a high level of scalability.

The main purpose of integrating Blockchain an edge computing technologies is in order to enhance efficiency and security within a distributed network. Additionally, both technologies are designed to work efficiently in large systems that are distributed. As a result, the two systems combine with ease. In the subsequent parts of this paper, we explore how both edge computing and blockchain technologies can be integrated.

IV. SYSTEM ARCHITCUTRE

*A. The Blockchain network*

The basis of the Blockchain network is to enable the signing of agreements among entities while at the same time increasing the security of the processes. Besides, it is recommended that the network efficiently function in terms of computation and be flexible enough to be easily managed by the implementers.

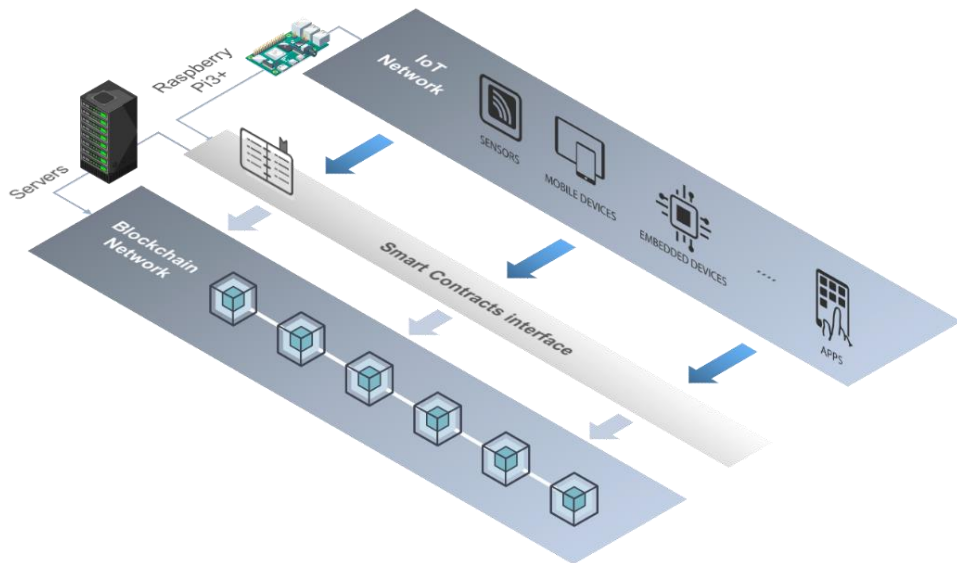

Fig. 2. Conceptual design for the system architecture

As a result, the use of the Ethereum network has been widely suggested. The fundamental reason being that the technology allows the configuration of location. Therefore, the manager of the system can easily determine the addresses that can be allowed into the network and which should not be given access. In addition to that, Ethereum technology allows the creation of new accounts that can be used in the Internet of Things devices with ease. In addition, Ethereum technology also supports the signing of agreements between entities in the network. Conclusively, Ethereum technology suffices as far as meeting the needs identified for the project is concerned.

*B. Smart contract*

This smart contract puts all the relevant security measures, the registration of information, and version management. As such, the system relies on the smart contract by and large. The functionalities of the contract are discussed later in this paper.

Solidity [14] is applied in the process of programming the smart contract. The programming language used is designed for creating smart contracts that can effectively function on Ethereum Virtual Machine. Besides, the language supports other areas that include inheritance, the data structures, and the mapping. The compiler of the language is known as SOLC. It comprises of various development tools that include Remix [15]. This is a platform for design, compilation, debugging, emulation as well as real-time testing of contracts.

*C. Servers*

Servers refer to blockchain network nodes that are involved in mining. This is to suggest that they complement IoT devices in the sense that they do what the devices would otherwise not do. The nodes have pre-installed Go Ethereum [16] or Geth, which enables them to affect the execution of the blockchain network. The servers have features that are greater than those of customers. In developing this project, we use a computer with the following features: 8GB RAM; 1TB; 7th Generation Intel I5 processor.

*D. Customers*

The entities in this architecture include IoT devices. They rely on Ethereum network servers to complement computational capabilities. These entities that are the customers make use of a program that is written in Java. The programming language increases the simulation of commercial IoT behavior by requesting resources available in the Blockchain network. The customers have Go Ethereum that is installed but which do not have the capabilities of carrying out mining operations.

It is worth noting that there are some IoT devices that are incapable of using such software. As a result, a proxy that acts as an intermediary is used. For the sake of this project, only the devices with the ability to use the software are considered. They use Raspberry Pi 3B+.

*E. The interface of Smart Contracts*

The application of Ethereum network technology is under the condition that customers at the edge be able to communicate with the network. This is possible and is done using the web3j interface[17]. Web3j is the reactive and also highly modular Java library. It facilitates the integration of nodes from the Ethereum network to the migration as well as the execution of the smart cards.

*F. Contract Migration*

Migrating the contract that is within the blockchain network helps in generating a useful code for execution. This migration process is enabled by Truffle[18], which is a free code for creating an environment that can sustain migration and a compilation. This is besides testing the smart cards that are in blockchain networks that use EVM.

## V. EXPERIMENTAL RESULTS

There are a number of experiments that are done to check the status of these functionalities: Version control and the downloading of specific data automatically; recording of an activity by system devices; detection of activities that are malicious and the resource distribution in the system.

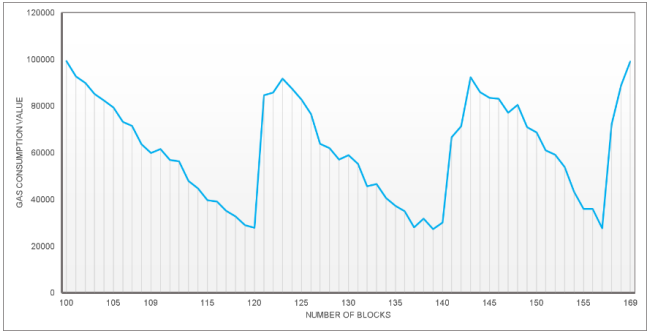

Fig. 3. Gas consumption of a single device.

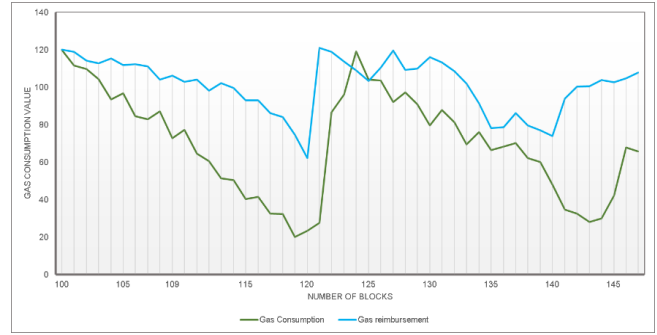

Fig. 4. Impact of a penalty on Gas reimbursement.

The initial test confirms the ability of a device to gather information from the blockchain network repository without an updated version of the software. To modify the software version, the migration of the contract is required and can only be executed when the administrator account verifies. With that, the IoT devices can download the latest updated version after rejection by the contract, which provides the URL for the download. The contract, therefore, is a gatekeeper to ensure that access to the network is only by the devices which have the required credentials.

Blockchain technology makes a note of all the transactions that take place in the network. Solidity comes in handy to store the data from the activities of the customer from each block in the chain. The other kind of data that is stored in the device data is collected from the regular transactions. In order for you to check how the operation of the device has been, you can retrieve the activities in the contract to verify the status of data that is stored already. We used Ether Gas, an economic token that is exchangeable, in measuring the levels of computational efforts required in the execution of experimental operations. The rationale behind using Gas is that it will prevent users with malicious intentions from disrupting the blockchain network by using programs that contain endless loop to jam the system.

The cost of each transaction in the contract migration is equivalent to the amount of Gas, which equates to Ethereum blockchain's fuel. The customer or entity that sends the Gas determines the amount that can be used and the price of unit gas given in ether units. If the gas limit given by the entity is surpassed, this leads to the cancellation of the transaction and payment to the minor.

Fig. 3 illustrates the resource distribution and the Gas amount owned by each device in the blockchain network. The transactions cost the resources of the customer, which is the sole device available in the blockchain system. When blocks have recovered the resources required, sharing of the resources is activated in an attempt to recover lost resources. Figure 4 gives the rationale of the system of penalizing the customers and Gas reimbursement. Here are transactions illustrated in the graph, showing the remitted payments. The blue consumer takes a greater transaction percentage than is allowed. The orange consumer, on the other hand, indicates the irregularities and the penalties on the blue consumer. Hence, it is at the time resources are being distributed that Gas is being reimbursed.

The blockchain network system is designed to function a bit differently during the initial stages of operations. The contract initializes the given variable the moment it is executed for the very first time. This is intended to regulate the intervals of blocks that are being formed. As a result, this leads to an increase in the cost of Gas of the specific transaction. Additionally, the initial access of the client to a smart contract is registered in the device map with the addition of customized data. Both Figures 3 and figure 4 illustrate instances where the system is stable.

## VI. CONCLUSION AND FUTURE WORK

We gather that further research is needed in order to enhance the delivery of data in distributed networks. This paper proposes a decentralized model developed from the Blockchain and edge computing technologies. Essentially, and as discussed above, edge computing is capable of enhancing the processing of data by bringing closer to the user the required computing services. This increases the speed of data processing and significantly reduces the costs of processing. This work has demonstrated the possibility of using blockchain technology in developing an architecture that could be useful in IoT devices and services on the network edge, while at the same time applying it within a reasonable computational cost. When the strengths of the blockchain technology are used together with those of edge computing, the resultant technology solves the issues of scalability and the security of the systems. As a future work, our interest will be in understanding proposed techniques on the frame base, the essence base, and the live content. In addition to that, we would like our future study to focus on the manual and the automated checks for Quality control. We could also look at the process of copying from the third party network.

ACKNOWLEDGEMENT

This work was supported by the Deanship of Scientific Research and Research Chair of Pervasive and Mobile Computing at King Saud University.